\documentclass[aps,prl,twocolumn,superscriptaddress,amsfont,graphicx,nofootinbib,preprintnumbers]{revtex4}%

\usepackage{graphicx}
\usepackage{float}
\usepackage{amsmath}
\usepackage{epsfig}

\newcommand{\be}{\begin{equation}}
\newcommand{\ee}{\end{equation}}
\newcommand{\ba}{\begin{eqnarray}}
\newcommand{\ea}{\end{eqnarray}}

\newcommand{\beq}{\begin{eqnarray}}
\newcommand{\eeq}{\end{eqnarray}}

\begin{document}
\begin{titlepage}

\begin{flushright}
\vbox{
\begin{tabular}{l}
    CERN-TH-2018-011,
    TTP18-004\\
\end{tabular}
}
\end{flushright}

\vspace{0.6cm}

\def\KIT{Institute for Theoretical Particle Physics, KIT, Karlsruhe, Germany}
\def\CERN{CERN Theory Division, CH-1211, Geneva 23, Switzerland}

\title{Higher order corrections to mixed QCD-EW contributions 
to Higgs production in gluon fusion
}

\author{Marco Bonetti }            
\email[Electronic address: ]{marco.bonetti@kit.edu}
\affiliation{\KIT}

\author{Kirill Melnikov}            
\email[Electronic address: ]{kirill.melnikov@kit.edu}
\affiliation{\KIT}

\author{Lorenzo Tancredi }            
\email[Electronic address: ]{lorenzo.tancredi@cern.ch}
\affiliation{\CERN}

\begin{abstract}
\vspace{2mm}
We present an estimate of the next-to-leading order QCD corrections 
to mixed  QCD-electroweak contribution to Higgs boson 
production cross section in gluon fusion, 
combining the recently computed three-loop virtual corrections 
and the approximate treatment of real emission in the soft approximation. 
We find that the NLO QCD corrections to mixed QCD-electroweak contributions 
are nearly identical to NLO QCD corrections to QCD Higgs production.  Our result  
confirms an earlier estimate of these ${\cal O}\left( \alpha\, \alpha_s^2 \right)$ 
effects  in~Ref.~\cite{Anastasiou:2008tj} 
and provides further support for the factorization approximation of QCD and electroweak 
corrections. 
\end{abstract}

\maketitle 

\end{titlepage}

Higgs boson production in gluon fusion is one of the central observables in Higgs physics at the LHC. This 
is because the majority of Higgs bosons are produced in this channel 
and also because the Higgs-gluon 
coupling is sensitive to heavy degrees of 
freedom that couple to gluons and receive their masses from the  Higgs mechanism. 

Given the importance of Higgs boson production in gluon fusion, in recent years 
its description by particle theorists   has been   provided  
with ever increasing accuracy.  The original computations of Higgs boson  production cross 
section in gluon fusion 
at leading \cite{Georgi:1977gs}, next-to-leading \cite{Dawson:1990zj,Graudenz:1992pv,Djouadi:1991tka} 
and next-to-next-to-leading order in perturbative QCD 
\cite{Harlander:2002wh,Anastasiou:2002yz,Ravindran:2003um} was 
recently extended to one order higher \cite{Anastasiou:2015ema}.
The residual uncertainty of the cross section related to uncalculated higher order QCD corrections 
was estimated to be of the order of two percent \cite{Anastasiou:2015ema,Anastasiou:2016cez}.   
To fully benefit from these remarkable achievements,  one 
needs to re-consider  the  many small contributions neglected in earlier 
calculations  and  study 
if they can change the gluon fusion cross section by a few percent.

A comprehensive analysis of the different contributions to Higgs boson gluon fusion cross section 
and their uncertainties was  recently 
presented  in Ref.~\cite{Anastasiou:2016cez}.  Among the 
uncertainties  are the top and bottom quark mass effects on the total cross section
in higher orders of perturbative QCD, the truncation of the 
expansion used to compute the  N$^3$LO contribution to the gluon fusion cross section, 
absence  of N$^3$LO parton distribution functions and the uncertainty in the value of NLO QCD corrections 
to the so-called mixed QCD-electroweak contribution to Higgs-gluon coupling.

In this paper we focus on higher-order QCD corrections to mixed QCD-electroweak contributions. 
These contributions appear at two loops for the first time  
and they are known to increase the 
leading order 
QCD cross section by about five percent \cite{Aglietti:2004nj,Actis:2008ug}. 
As it is often the case in Higgs physics,  it is not 
clear how this result changes when higher order QCD 
corrections to  the gluon fusion cross section are accounted for. Indeed, 
since  the NLO QCD 
corrections to top-mediated Higgs production in gluon fusion are close to ${\cal O}(100\%)$,
it  is important to know if  these large  corrections also apply to mixed QCD-electroweak contributions 
since, depending  on  whether they do or they  do not, 
the cross section changes by an amount that is not negligible 
at the level of the precision target of a few percent.
 
It is difficult to compute the NLO QCD corrections to mixed QCD-electroweak contribution. Indeed, 
this contribution  appears at two loops for the first time, so that the computation of NLO QCD corrections 
to it requires the calculation of three-loop Feynman diagrams to account for  virtual corrections and 
two-loop four-point functions to evaluate the real emission corrections.  Both of these tasks 
are quite formidable. 

To overcome this difficulty, in Ref.~\cite{Anastasiou:2008tj} the NLO QCD 
 corrections to mixed QCD-EW contributions were computed in an unphysical 
limit where the masses of electroweak gauge bosons are 
considered to be significantly larger than the Higgs boson mass.
For such mass  hierarchy  one can perform a systematic large mass expansion of the corresponding 
Feynman graphs \cite{Smirnov:2013} that, effectively, 
turns the QCD-electroweak contribution to Higgs gluon
coupling to a point-like interaction vertex.  
It is clear that the assumed mass hierarchy is questionable and that 
the result can only be considered as an {\it estimate} 
of the NLO QCD corrections to QCD-electroweak contribution. 

According to Ref.~\cite{Anastasiou:2016cez},  at NLO QCD, the QCD-electroweak contributions  
increase the 
gluon  fusion cross section 
by about $5 \pm 1$ percent.  The uncertainty estimate shown 
here refers to an attempt to quantify a possible error 
caused by the  unphysical approximation for Higgs  and vector 
boson masses employed in Ref.~\cite{Anastasiou:2008tj}. 

To improve on this result, one has to compute the NLO QCD corrections 
to mixed QCD-electroweak contributions to Higgs boson production 
cross section in gluon fusion for the correct relation between the Higgs 
boson and the electroweak gauge boson masses. 
Recently, we made the first step in this direction 
by calculating the relevant  three-loop virtual corrections \cite{Bonetti:2017ovy}.
To obtain the corrections to gluon fusion cross section, one needs to combine this 
result with the real emission contributions that involve two-loop four-point functions 
with several mass scales; computing them is quite  complicated. 
While work on these real emission contributions is in progress, the computation 
of the virtual corrections reported in Ref.~\cite{Bonetti:2017ovy} opens up a way 
to provide an estimate of the NLO QCD effects to mixed QCD-electroweak 
contributions that is {\it conceptually  different} 
from what has been done   in Ref.~\cite{Anastasiou:2008tj}. 
As such, it will either provide additional support for the size of 
mixed QCD-electroweak 
contributions estimated in  Ref.~\cite{Anastasiou:2008tj} or it 
will indicate the  potential deficiencies of such an estimate.  
Either of these outcomes is important for  
understanding the current theoretical precision 
on the Higgs boson production cross section in gluon fusion. 

Our estimate of the NLO QCD corrections to Higgs boson production in gluon fusion 
is based on an observation that QCD corrections to this process 
can be relatively well described by the soft-gluon approximation 
\cite{Catani:2001ic,deFlorian:2012za,Ball:2013bra}. The soft gluon approximation 
accounts for  contributions of real gluon emissions by a universal formula that depends 
on leading order cross section. The only non-universal piece that 
needs to be provided are the virtual corrections computed by us recently~\cite{Bonetti:2017ovy}.

We now explain the details of the calculation. 
The Higgs boson production cross section in gluon fusion can be written as 
\be
\sigma = \int \limits {\rm d} x_1 {\rm d} x_2 g(x_1,\mu) g(x_2,\mu) \left ( z \sigma_0 \right ) 
G(z,\mu,\alpha_s), 
\label{eq1}
\ee
where $z = m_H^2/(sx_1 x_2)$, $m_H$ is the mass of the Higgs boson,  $s$ is the 
center-of-mass energy squared of the hadronic collision, $\alpha_s \equiv \alpha_s(\mu)$ is the strong 
coupling constant and $\mu$ denotes factorization 
and renormalization scales that we set equal to each other.  Note that the only partonic channel that 
contributes in the soft approximation is the $gg$-channel. 

The leading order cross section $\sigma_0$ reads 
\be
\sigma_0 = \frac{\alpha_s^2}{576 \pi v^2} F_0(m_H,m_W,m_Z),
\label{eq2}
\ee
where the form factor $F_0$ contains   QCD and mixed QCD-electroweak contributions
at leading order. Finally, at leading order
$$G(z,\mu, \alpha_s) = \delta(1-z)\,.$$
To evaluate $F_0$, we  use the following numerical values 
for Standard Model parameters $m_H = 125~{\rm GeV}$, $m_{W} = 80.398~{\rm GeV}$, 
$m_Z = 91.88~{\rm GeV}$, $\alpha_{\rm QED} = 1/128.0$, $\sin^2 \theta_W = 0.2233$, 
$G_F = 1.16639 \times 10^{-5}/{\rm GeV}^2$. We also use the Higgs field vacuum 
expectation value defined as $v= ( G_F \sqrt{2} )^{-1/2}$. We employ numerical 
values for  $\alpha_s$ and 
gluon parton distribution functions as provided by the NNPDF30 set \cite{Ball:2014uwa}. 
Specifically, we use ${\rm NNPDF}30_{\rm lo-as-0130}$ 
and ${\rm NNPDF}30_{\rm nlo-as-0118}$ for leading and next-to-leading 
order computations, respectively. 

The  leading order cross section Eq.(\ref{eq2}) is normalized in such 
a way that $F_0 = 1$  if only   pure QCD contributions  to the form factor $F$ are  taken into account.
Including also the QCD-electroweak contribution, the result reads 
\be
F_0 = |A_0|^2,
\label{eq3}
\ee
where 
\be
A_0 = 1 - \frac{3 \alpha^2 v^2}{ 32 m_H^2 \sin^4 \theta_W} \left ( C_W A_W + C_Z A_Z \right ),
\ee
with 
\be
\begin{split}
& C_W = 4, \\
&     C_Z = \frac{2}{\cos^4\theta_W} \left ( \frac{5}{4} 
     - \frac{7}{3} \sin^2 \theta_W + \frac{22}{9}\sin^4 \theta_W \right ),
\end{split} 
\ee
and 
\be
\begin{split}
& A_W = -10.71693 - \mathrm{i} \;2.302953, \\
& A_Z =  -6.880846 - \mathrm{i}\; 0.5784119.
\end{split}
\ee
These numerical values for the mixed QCD-EW amplitudes at leading order  
follow from analytic  calculations reported in \cite{Aglietti:2004nj,Bonetti:2016brm}. 
To obtain $A_{W,Z}$, we consistently neglect the top quark contributions in 
case of $Z$-exchange  amplitudes and the third generation contribution in case of 
$W$-exchange amplitudes;\footnote{ Top quark contributions to mixed QCD-electroweak 
corrections are known to be tiny \cite{Actis:2008ug}.}
 we do exactly the same when we compute NLO QCD corrections 
to QCD-electroweak contributions as described below. 
We note that, according to Eq.(\ref{eq3}), we include the square of the mixed QCD-electroweak 
contribution to the cross section. Numerically, this makes a tiny difference and we do it 
for the sake of convenience. 

As explained earlier,  to extend this result beyond leading order, we use the soft-gluon approximation 
to describe the real emission corrections. The corrections to the function $G(z,\mu,\alpha_s)$ then follow 
from the 
soft approximation to the $gg \to Hg$ matrix element squared where, independent of the 
hard process, the gluon emission is described by an eikonal factor.  
Integrating the eikonal factor over the gluon phase space and removing the collinear singularities 
by renormalization of the parton distribution functions, one finds \cite{Catani:2001ic,deFlorian:2012za}
\be
\begin{split}
& G(z,\mu,\alpha_s) = \delta(1-z) 
\\
& + \frac{\alpha_s}{2 \pi} \left [ 
8 C_A \left ( D_1(z) + \frac{D_0(z)}{2}  \ln \frac{m_H^2}{\mu^2} \right ) 
\right. \\
& \left.   + \left ( \frac{2 \pi^2}{3} C_A + V \right ) \delta(1-z) \right ]. 
\end{split}
\label{eq7}
\ee
Here, $C_A=3$ is the number of colors and $D_0=[1/(1-z)]_+$, $D_1 = [\ln(1-z)/(1-z)]_+$ are plus distributions. $V$ is the ratio of the infra-red 
subtracted virtual corrections to the leading order cross section. 
Note that this quantity $V$  represent {\it the only non-universal contribution  
in the soft limit}, which means that it is this quantity that may, potentially, change the relative 
size of electroweak corrections to Higgs production cross section at leading and next-to-leading 
orders in perturbative QCD.   The infra-red subtracted  virtual corrections are obtained 
from the results for NLO QCD corrections to mixed QCD-electroweak contributions 
reported in Ref.~\cite{Bonetti:2017ovy} and from the known NLO QCD corrections to 
leading order production cross section \cite{Dawson:1990zj}. 
We write 
\be
V =2  {\rm Re}( A_{1,\rm fin} A_0^*)/ |A_0|^2,
\ee
where 
\be
A_{1,\rm fin} = 
\frac{11}{2} - 
\frac{3 \alpha^2 v^2}{ 32 m_H^2 \sin^4 \theta_W}
\left ( C_W A_W^{(1)} + C_Z A_Z^{(1)} \right ),
\ee
and \cite{Bonetti:2017ovy}
\be
\begin{split}
& A_W^{(1)} = -11.315691 - \mathrm{i}\; 54.029527,
\\  
& A_Z^{(1)} = -2.975666 - \mathrm{i}\; 41.195540. 
\end{split}
\ee

In principle, the above results allow us to compute the Higgs boson cross section in the soft 
gluon approximation. However, it is known that the soft gluon approximation underestimates the NLO  
corrections. An attempt to improve on this by constructing subleading terms was undertaken 
in Ref.~\cite{Ball:2013bra}.  It was argued there,  using 
analiticity considerations  in Mellin space  and information on 
universal subleading terms in the $ z \to 1$  limit that arise from 
soft-gluon kinematics and, also, from the collinear splitting kernels, 
that  a useful  extension   of the soft approximation is obtained 
by replacing  the plus-distribution $D_1(z)$
that appears in Eq.(\ref{eq7})  with 
\be
D_1(z) \to D_1(z) + \delta D_1(z),
\ee
where 
\be
\delta D_1(z) = (2-3z + 2 z^2) \frac{\ln( (1-z)/\sqrt{z})}{1-z} - \frac{\ln(1-z)}{1-z}.
\label{eq12}
\ee
Note that $\delta D_1(z)$ is an integrable function of $z$ and not a plus-distribution.

It is now straightforward to use the above results  to estimate the NLO QCD corrections 
to mixed QCD-electroweak contribution  in the soft gluon approximation. We take $s = (13\,\textup{TeV})^2$. We use 
NNPDF30 sets \cite{Ball:2014uwa} to compute the gluon fusion cross section and we use LO and NLO 
parton distribution functions to perform computations in respective perturbative 
 orders.  We set the values of the 
factorization and the renormalization scales equal to each other. The central 
value for both scales is taken  to be $\mu = m_H/2$.  We note, however, 
that  our main result -- the relative change 
in QCD cross sections due to  mixed QCD-EW contributions -- 
is practically independent of the central scale.  Computing the Higgs production 
cross section using Eq.~(\ref{eq1}), we obtain the following results 
\be
\begin{split}
& \sigma_{\rm QCD}^{\rm LO} = 20.6~{\rm pb},\;\;\;\;  \sigma_{\rm QCD/EW}^{\rm LO} = 21.7~{\rm pb},
\\
& \sigma_{\rm QCD}^{\rm NLO} = 32.66~{\rm pb},\;\;\;\; \sigma_{\rm QCD/EW}^{\rm NLO} = 34.41~{\rm pb}.
\end{split} 
\label{eq13}
\ee
It follows from these numbers that the electroweak-QCD contributions increase 
{\it both} the LO and NLO  cross sections by  $5.3-5.5$ percent. 
This result is consistent with the estimate of the impact of mixed QCD-EW corrections 
obtained in   Ref.~\cite{Anastasiou:2008tj}.  

As a check on the robustness of this  result, we repeat the same 
computation setting $\delta D_1(z)$ in Eq.(\ref{eq12}) 
to zero. 
Since, as we explained earlier,  by introducing  $\delta D_1(z)$ 
we attempt to describe radiation beyond the soft limit, by  removing it from the 
computation  we check the sensitivity of the result to the part of computation that we poorly control. We 
find ($\delta D_1 \to 0$)
\be
 \sigma_{\rm QCD}^{\rm NLO} = 26.30~{\rm pb},\;\;\;\; \sigma_{\rm QCD/EW}^{\rm NLO},
= 27.70~{\rm pb}. 
\label{eq14}
\ee
It follows that also in this case the mixed QCD-electroweak contribution e
xceeds the QCD cross section 
by about $5.35$ percent. 

Finally, we can also check what happens if we use the {\it exact} NLO  results 
for QCD contributions,  and only employ the  soft approximation to 
describe the mixed QCD-EW contribution.  The corresponding NLO QCD cross section 
can be obtained with {\tt MCFM }\cite{mcfm}.  For $\mu = m_H/2$, the result  
reads\footnote{All partonic channels are now included.}
$\sigma_{\rm QCD}^{\rm NLO, full}  = 35.4~{\rm pb}$.  The change in NLO QCD 
cross section caused by  QCD-EW contributions is 
obtained from Eqs.(\ref{eq13},\ref{eq14}). 
We find $\delta \sigma^{\rm NLO}_{\rm QCD-EW} = 1.6 - 
2~{\rm pb}$, depending on whether we include improved or unimproved 
soft approximation. Computing 
the ratio $\delta \sigma^{\rm NLO}_{\rm QCD-EW}/ \sigma_{\rm NLO}^{\rm QCD,full}$, we obtain 
$(4.7 - 5.5) \times 10^{-2}$, consistent with other  estimates described above.

The soft approximation for real gluon emission that we employ here does not describe 
correctly the structure-dependent radiation that arises when  gluons are emitted from the 
``interior'' of the loop amplitude. However, the contribution 
of the true structure-dependent radiation to the cross section is 
suppressed by {\it two powers} of the gluon energy relative to the soft gluon
approximation \cite{subl}.  For this reason, there is a good chance that the 
structure-dependent radiation plays a relatively minor role and that the 
soft gluon approximation employed by us in this paper provides sufficiently 
good description of real emission. 

To conclude, we employed the soft-gluon approximation and the recent computation of three-loop 
virtual corrections in Ref.~\cite{Bonetti:2017ovy} to estimate the size of 
the NLO QCD corrections to  mixed QCD-electroweak contributions to gluon fusion cross section. 
We find that mixed QCD-electroweak contributions increase both the leading and next-to-leading 
order cross sections by $5.4$ percent. This result is consistent 
with an estimate of this corrections presented in \cite{Anastasiou:2008tj}.  
Further improvements are only possible if the real emission contributions are computed exactly. 
This is a very challenging problem that, hopefully, can be solved using the many recent advances 
in the technology of loop computations.

\vspace{0.5cm}
{\bf \hspace{-0.6cm} Acknowledgements} 
We would like to thank Fabrizio Caola for useful comments. 
We are grateful to C.~Anastasiou for pointing out a mistake in the previous 
version of this paper. 
The research of K.M. was supported by the German Federal Ministry for %
Education and Research (BMBF) under %
grant 05H15VKCCA.
The work of M.B. was supported by a graduate fellowship 
from DFG Research Training Group 1694/2 ``Elementary particle physics 
at highest energy  and precision''.  The research 
of L.T. was supported by the ERC starting grant 637019 ``MathAm''.

\end{document}